\documentclass[prb,11pt,showkeys,showpacs]{revtex4-1}

\usepackage[latin1]{inputenc} 
\usepackage[T1]{fontenc} 
\usepackage[english]{babel}  

\usepackage{amsmath}    
\usepackage{graphicx}   
\usepackage{verbatim}   
\usepackage{color}      
\usepackage{subfigure}  
\usepackage{hyperref}   
\begin{document}

\title{Implementation of transmission functions for an optimized three-terminal quantum dot heat engine}
\author{Christian H. Schiegg, Michael Dzierzawa, and Ulrich Eckern}
\affiliation{Institute of Physics, University of Augsburg, 86159 Augsburg, Germany}

%\date{\today}

\begin{abstract}
We consider two modifications of a recently proposed three-terminal quantum dot heat engine. 
First, we investigate the necessity of the thermalization assumption, namely that electrons are always thermalized by inelastic processes when traveling across the cavity where the heat is supplied.
Second, we analyze various arrangements of tunneling-coupled quantum dots in order to implement a transmission function that is superior to the Lorentzian transmission function of a single quantum dot.
We show that the maximum power of the heat engine can be improved by about a factor of two, even for a small number of dots, by choosing an optimal structure.
\end{abstract}
%\pacs{73.63.-b, 05.70.Ln, 72.15.Jf, 84.60.Rb}
\maketitle

\section{Introduction}
\label{introduction}
\begin{figure}[h!]
  \includegraphics[width=0.4\textwidth]{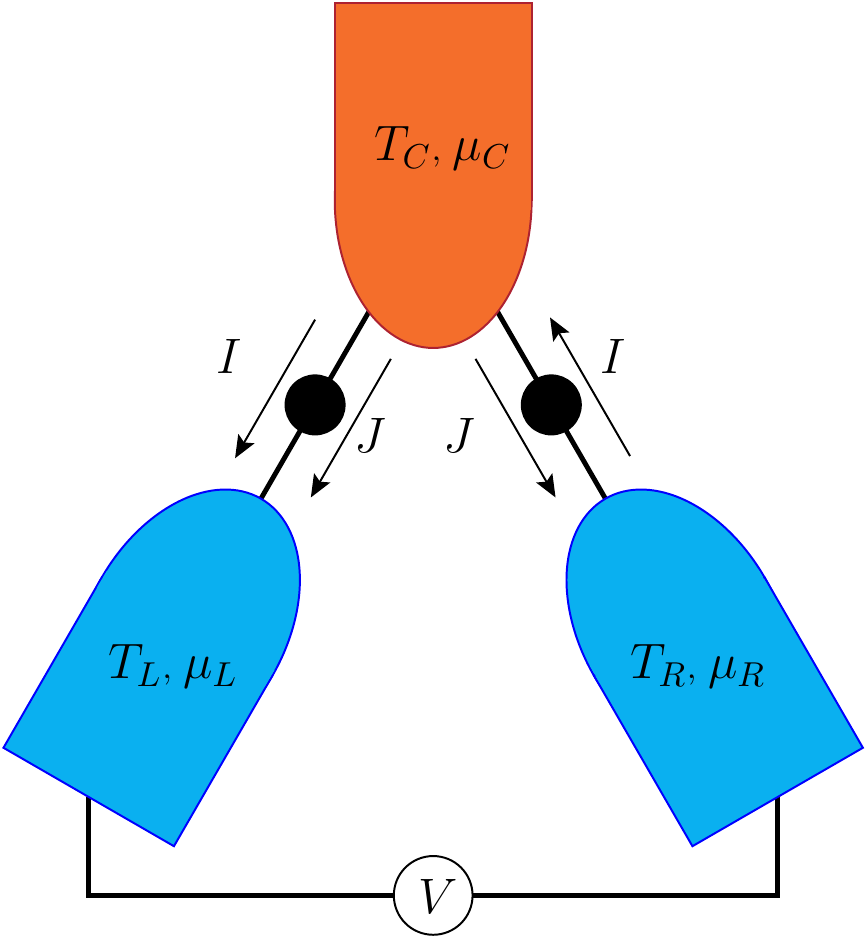}
	\caption{Three-terminal quantum dot heat engine proposed in Ref.~\onlinecite{jordan}.
	The cold leads (blue) are connected with the hot cavity (red) via quantum dots (black bullets). 
	The direction of heat currents $J$ and charge currents $I$ is indicated by arrows, and $V$ is the generated voltage at an external load.
	Note that for $e < 0$ the charge current flows in opposite direction to the particle current.
	}
	\label{model_heat_engine}
\end{figure}

In recent years there has been increasing interest in the development of nanoscale devices that are able to convert heat into electricity.\citep{sothmann,dubi,mahan}
The functionality of such devices generally depends on the combined transport of charge and heat through quantum systems coupled 
to macroscopic reservoirs.
Various layouts for such heat engines have been discussed, most of them based on quantum dots.\citep{jordan,hershfield,beenakker,humphrey,broeck,nakpathomkun,esposito,sothmann2,sanchez,koch,mani,hicks,yamamoto,choi,thierschmann,zhang}
In particular, for the heat engine proposed in Ref.~\onlinecite{jordan} two quantum dots with a single energy-level are used as energy filters in order to generate a directed charge current.
A sketch of this three-terminal heat engine is shown in Fig.~\ref{model_heat_engine}. 
It consists of two leads, left and right, that are in thermal equilibrium at temperature $T_L = T_R $, and a central reservoir (cavity) at a higher temperature $T_C$.
Electrons from the left lead can move into the cavity by tunneling through the first quantum dot, and from there into the right lead via the second quantum dot. 
A particle current flowing from the left lead into the cavity is achieved by shifting the energy level of the connecting quantum dot downwards such that the 
thermal occupation (Fermi function) close to the level position is higher on the left side due to the temperature difference.
Shifting the energy level of the second quantum dot upwards by the same amount yields an equal current from the cavity into the right lead, such that the  net charge current 
into the cavity vanishes. 
On the other hand, due to the temperature difference there is a positive heat current from the cavity into the leads; therefore, in order to achieve a stationary state, a constant supply of heat into the cavity is required to compensate the loss. 
Connecting left and right leads by an external load allows to extract electrical power $P = IV$ from the engine, where the voltage $V$ adjusts itself according to the resistance of the load. 

In the original proposal \citep{jordan} it is assumed that electrons entering the cavity from the left lead stay there long enough in order to be thermalized by inelastic processes before they proceed to the right lead.
Coherent tunneling from the left to the right lead via the cavity is therefore explicitly excluded. 
This assumption has the technical advantage that the three-terminal problem of Fig.~\ref{model_heat_engine} can be treated as two 
effectively independent two-terminal systems, tunneling from L to C and tunneling from C to R, respectively. 
However, one may ask the question whether this assumption is essential for the proper functionality of the heat engine, and what happens if the setup of the engine is modified such that coherent tunneling from the left to the right lead is enabled. 
In Sec.~\ref{sec_coh_vs_incoh} we will address this question by considering two non-interacting tight-binding models that represent the two aforementioned cases. 

The performance of a heat engine is usually characterized by the maximum power that can be generated, and by its efficiency, i.e., the ratio between the electrical power and the heat current. 
For non-interacting quantum systems the charge current $I$ and the heat current $J$ can be calculated using the Landauer formula,\citep{landauer,datta} e.g., for the currents between $L$ and $C$:
\begin{equation}
I = \frac{2e}{h}\int dE\, T(E) \big[f_L(E)-f_C(E)\big]
,
\label{landauer_charge}
\end{equation}
\begin{equation}
J=\frac{2}{h}\int dE\, T(E)(E-\mu_L) \big[f_L(E)-f_C(E)\big]
,
\label{landauer_heat}
\end{equation}
where $f_L$ and $f_C$ are the Fermi functions for the left lead with temperature $T_L$ and chemical potential $\mu_L= - eV/2$, and for the cavity with temperature $T_C$ and chemical potential $\mu_C = 0$, respectively. 
The factor $2$ is due to the spin, and $e < 0$ denotes the charge of an electron.
Besides the parameters temperature and voltage entering the Fermi functions, both currents depend only on the transmission function $T(E)$. 
It is therefore natural to ask which is the optimal transmission function, based on certain criteria for the engine's performance, e.g., efficiency at maximum power. 
From the properties of the Landauer integrals (\ref{landauer_charge}) and (\ref{landauer_heat}) it has been argued \citep{whitney2014,whitney2015} that the optimal transmission function is rectangular-shaped with $T(E) = 1$ inside a certain energy window, and zero outside.
A second and less obvious question is how this ideal transmission function can be practically implemented using tunnel-coupled quantum dots as building blocks. 
We will address this question in Sec.~\ref{sec_opt_trans}, where we calculate and compare the transmission functions of various arrays of coupled quantum dots.
In particular, we present closed analytical expressions for homogeneous linear chains and ring structures of arbitrary size, and derive the hopping parameters for an optimized inhomogeneous linear chain of up to 13 quantum dots.
A brief summary is given in Sec.~\ref{sec_summary}.

\section{Coherent vs. incoherent tunneling}\label{coherent_vs_incoherent}
\label{sec_coh_vs_incoh}

In this section we present model calculations for nanoscale heat engines, with emphasis on the comparison of the incoherent with the coherent heat engine as discussed above.
In our calculations we model both the leads and the central cavity as infinitely long non-interacting one-dimensional tight-binding systems with hopping parameter $t_0$ and dispersion $E = -2t_0 \cos k$; the lattice constant is set to unity. 
For simplicity, we restrict ourselves to the wide-band limit $t_0 \rightarrow \infty$, i.e., $k\rightarrow \pi/2$, such that the density of states in the leads can be considered as constant.  

\subsection{Incoherent heat engine}\label{sub_incoherent}
\label{section_incoherent}
\begin{figure}[ht]
  \includegraphics[width=0.6\textwidth]{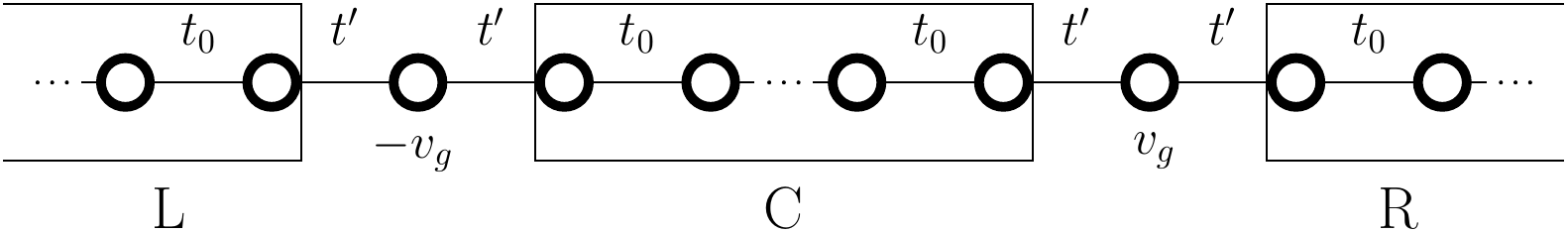}
	\caption{Tight-binding model for the incoherent heat engine}
	\label{model_incoherent}
\end{figure}
\begin{figure}[ht]
  \includegraphics[width=0.47\textwidth]{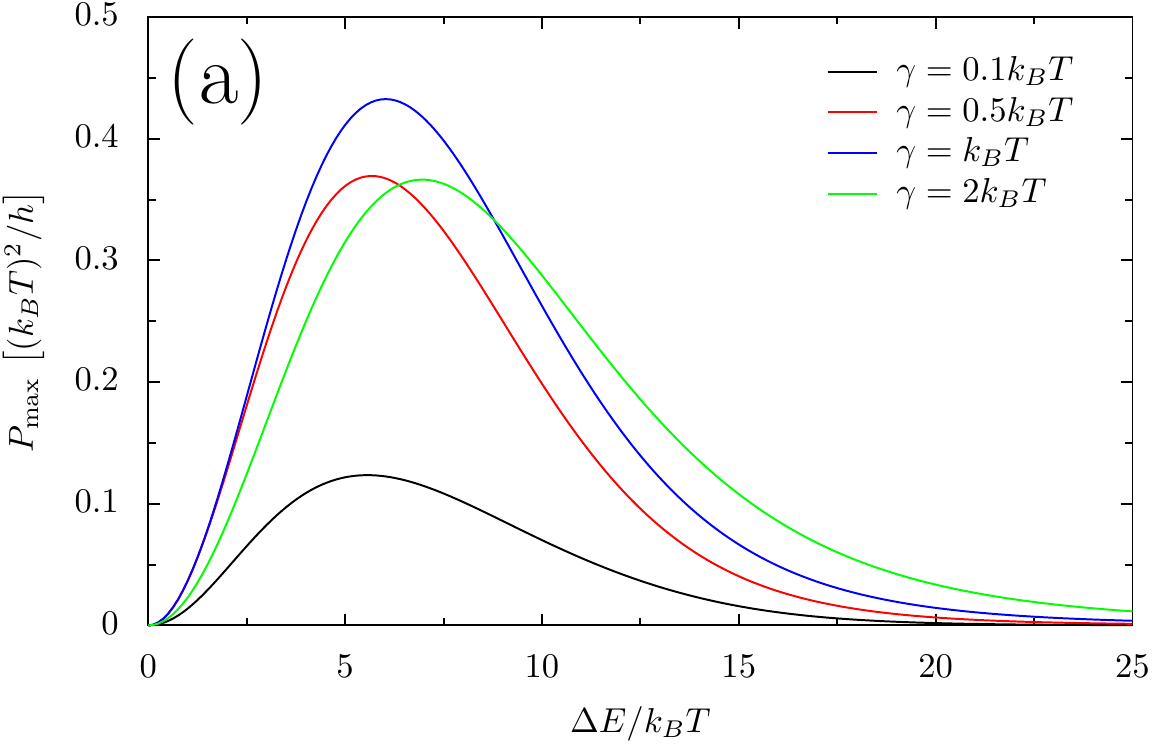}
  \includegraphics[width=0.47\textwidth]{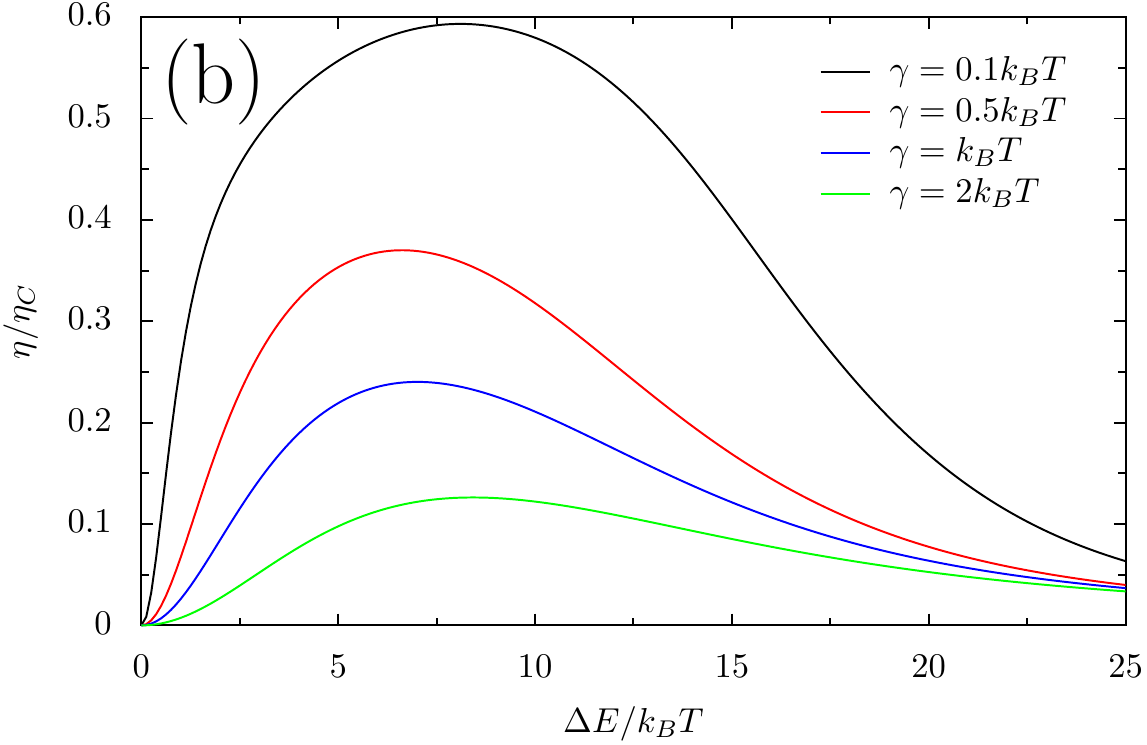}
	\caption{Maximum power and efficiency at maximum power of the incoherent heat engine as function of $\Delta E$ for 
	$\Delta T/T=1$ and different values of $\gamma$. The efficiency $\eta$
	is normalized by the Carnot efficiency, $\eta_C=1 - T_L/T_C$.}
	\label{power_eff_1_dot_inco}
\end{figure}

As displayed in Fig.~\ref{model_incoherent}, the energy levels of the quantum dots connecting left and right leads with the central region are antisymmetrically shifted by local potentials $\mp v_g$, and the hopping parameter between the quantum dots and the reservoirs is $t'$ on both sides. 
The corresponding Hamiltonian reads
\begin{equation}
\hat{H}=-\sum_{i,\sigma}t_i(\hat{c}^{\dagger}_{i+1,\sigma}\hat{c}_{i,\sigma}^{\phantom{\dagger}}+\mathrm{h.c.})+\sum_{i}v_i\hat{n}_i,
\end{equation}
where the hopping amplitudes $t_i$ and local potentials  $v_i$ are indicated in Fig.~\ref{model_incoherent}.
The operators $\hat{c}_{i,\sigma}^{\phantom{\dagger}}$ ($\hat{c}^{\dagger}_{i,\sigma}$) denote the usual fermion annihilation (creation) operators, and $\hat{n}_i=\sum_{\sigma}\hat{c}^{\dagger}_{i,\sigma}\hat{c}_{i,\sigma}^{\phantom{\dagger}}$.
The dots inside the central region indicate that it is of infinite size, and can be regarded as a particle and energy reservoir at fixed temperature $T_C$ and chemical potential $\mu_C = 0$.
In order to satisfy the constraint of vanishing net charge current into the central region one has to choose the chemical potentials in the leads antisymmetrically, $\mu_{L,R} = \mp eV/2$. 
Then the charge currents are given by $I_{L\rightarrow C}=I_{C\rightarrow R}=I$, and the heat currents by $J_{C\rightarrow L}=J_{C\rightarrow R}=J$.
Therefore, it is sufficient to calculate the charge and heat current between the left lead and the central reservoir. 
In the following, we parametrize the temperatures of the reservoirs as $T_L = T_R = T - \Delta T/2$ and $T_C = T + \Delta T /2$.
In the wide-band limit, the transmission function of the left quantum dot (with potential $-v_g$) is given by the Lorentzian
\begin{equation}
T(E)=\frac{\gamma^2}{\gamma^2+(E+v_g)^2},
\end{equation}
where $\gamma=2t'^2/t_0$.  
Using the Landauer formulas (\ref{landauer_charge}) and (\ref{landauer_heat}) it is straightforward to numerically calculate the currents, and from there the power and efficiency for given parameters $T, \Delta T$, $\gamma$, and energy difference between the two quantum dots, $\Delta E = 2v_g$, as a function of the voltage $V$. 
The maximum power $P_{\mathrm{max}}$ (with respect to $V$) and the efficiency at maximum power are shown in Fig.~\ref{power_eff_1_dot_inco} versus $\Delta E$ for various values of $\gamma$. 
All energies are expressed in units of $k_B T$, and we set $\Delta T = T$.
The overall maximum power is reached for $\gamma \approx k_B T$ and $\Delta E \approx 6 k_B T$, in agreement with the results given in 
Refs.~\onlinecite{sothmann,jordan}.
The data presented in Fig.~\ref{power_eff_1_dot_inco} will be used in Sec.~\ref{sec_opt_trans} as benchmark results for the comparison with the performance parameters of more efficient transmission functions.  

\subsection{Coherent heat engine}\label{sub_coherent}

\begin{figure}[ht]
  \includegraphics[width=0.4\textwidth]{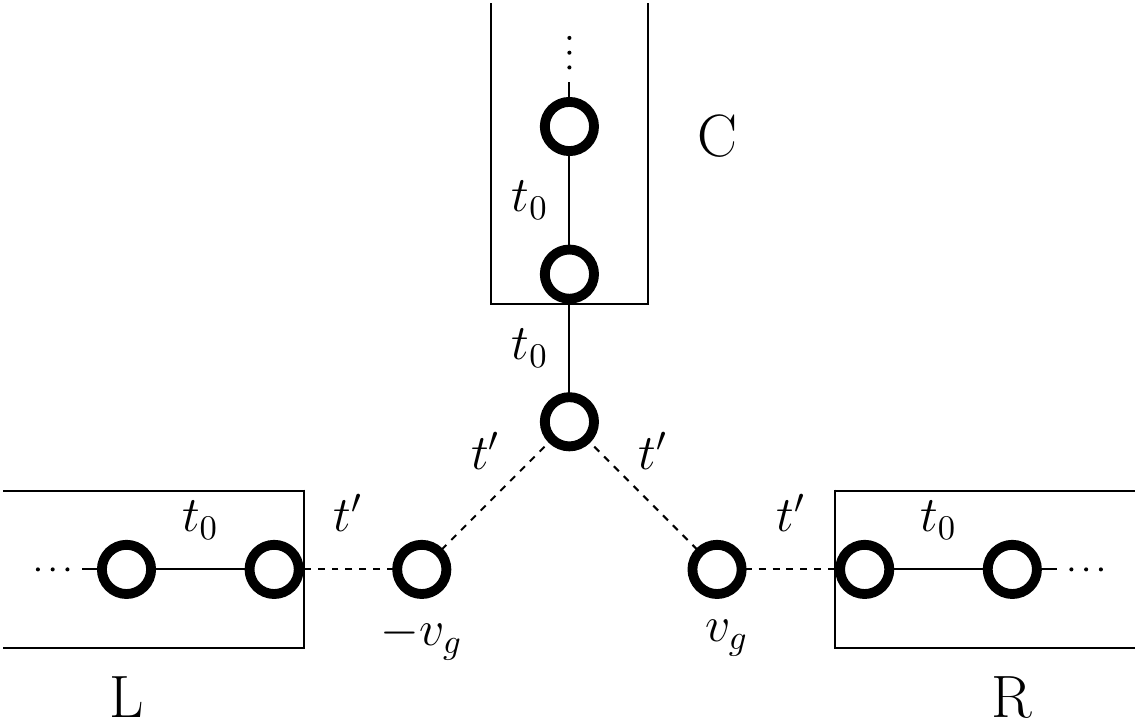}
	\caption{Tight-binding model for the coherent heat engine}
	\label{model_coherent}
\end{figure}

\begin{figure}[ht]
  \includegraphics[width=0.47\textwidth]{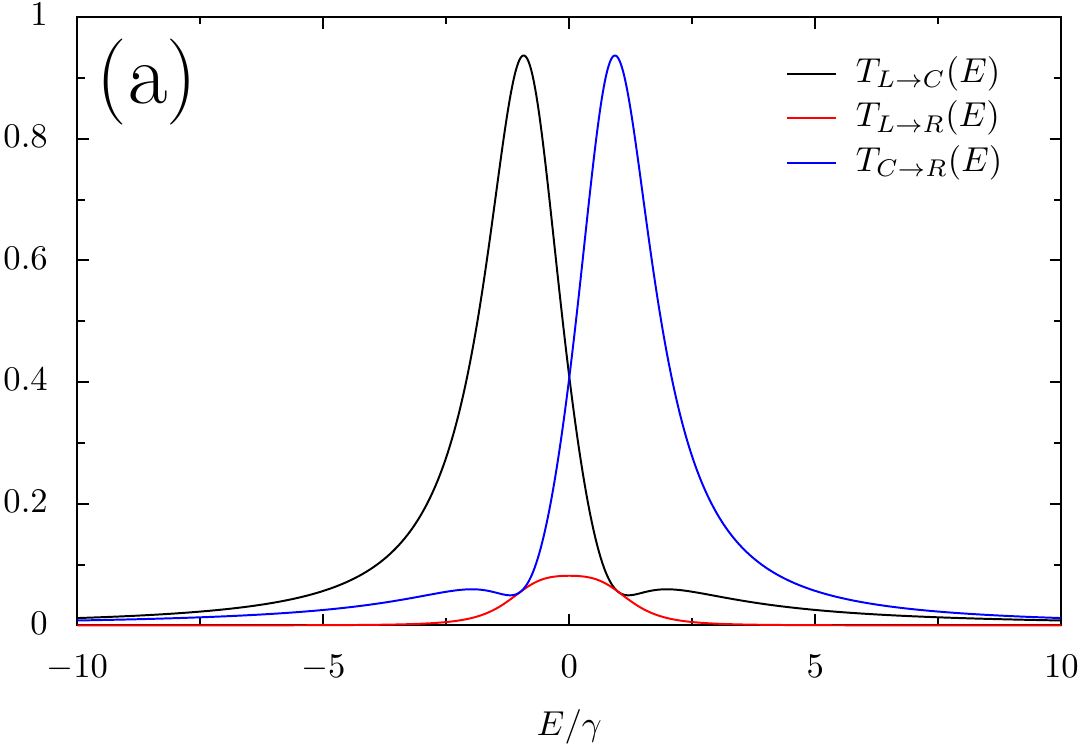}
  \includegraphics[width=0.47\textwidth]{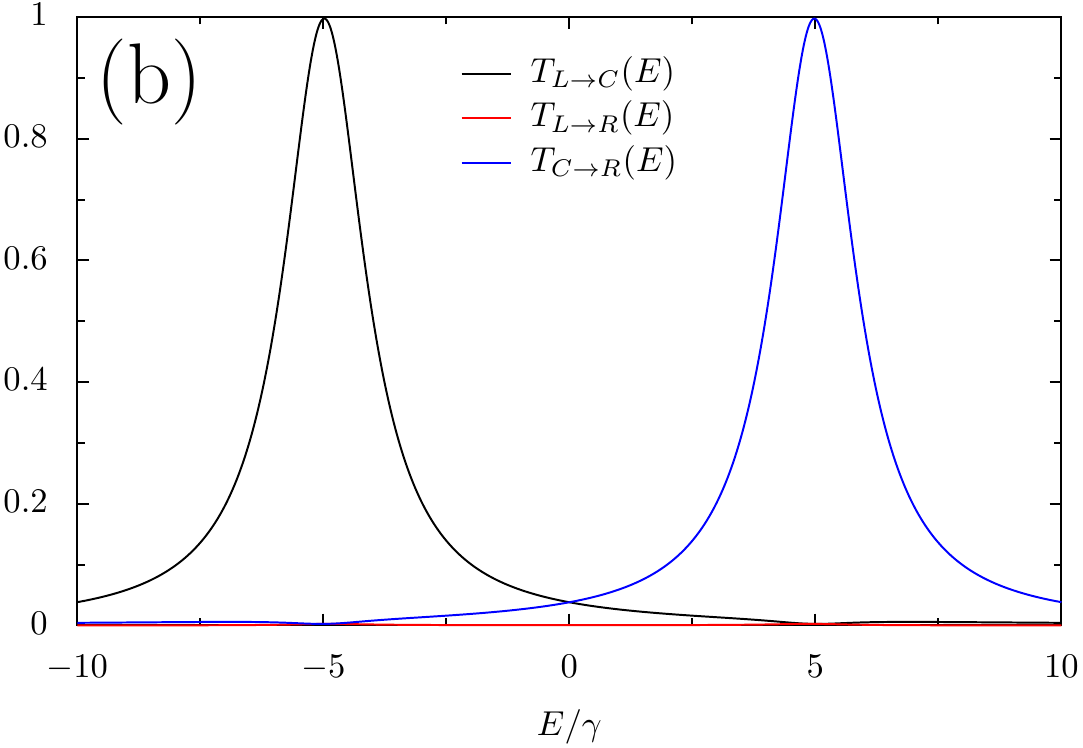}
	\caption{Transmission functions of the coherent three-terminal model (see Fig.~\ref{model_coherent}) for $\Delta E=2\gamma$ (a) and $\Delta E = 10\gamma$ (b).}
	\label{coherent_trans}
\end{figure}

Figure \ref{model_coherent}  depicts the tight-binding model for the coherent version of the heat engine. 
In contrast to the previous case, the central reservoir $C$, which supplies the heat, is connected to left and right leads via a single site, such that electrons can coherently tunnel from the left to the right lead without being thermalized.
As a consequence, we have to use the three-terminal generalization of the Landauer formula, e.g., the total charge current out of the left lead is given by
\begin{equation}
I_L=\frac{2e}{h}\sum_{\alpha=C,R}\int \mathrm{d}E\,  \mathrm{T}_{L\rightarrow \alpha}(E)\big[f_L(E)-f_\alpha(E)\big]
\label{landauer_charge_coherent}
.
\end{equation}
In order to determine the transmission functions $\mathrm{T}_{L\rightarrow C}$ and $\mathrm{T}_{L\rightarrow R}$, we solve the discrete Schr\"odinger equation $\hat{H}|\phi\rangle=E|\phi\rangle$ with an ansatz representing an incoming plane wave originating from the left lead with wavenumber $k$: 
\begin{equation}
\phi_n=
\left\{
\begin{array}{ll}
\mathrm{e}^{\mathrm{i}kn}+b\, \mathrm{e}^{-\mathrm{i}kn}\ \ \ & \mathrm{for}\ n\ \in\ L ,\\
c \, \mathrm{e}^{\mathrm{i}kn} &\mathrm{for}\ n\ \in\ C, \\
d \, \mathrm{e}^{\mathrm{i}kn} &\mathrm{for}\ n\ \in\ R. \\
\end{array}
\right.
\label{plane_wave}
\end{equation}
The transmission functions are given by $T_{L\rightarrow C}=|c|^2$ and $T_{L\rightarrow R}=|d|^2$.
Inserting this ansatz into the Schr\"odinger equation in the wide-band limit ($k\rightarrow \pi/2$) yields the linear system
\begin{equation}
\left(
\begin{array}{ccccc}
-\mathrm{i}t_0 & -t'& 0 & 0 & 0 \\
 -t'&-(E+v_g)& -t' & 0 & 0 \\
0 & -t' & -\mathrm{i}t_0 & -t'& 0 \\
0 & 0 & -t'&-(E-v_g)& -t' \\
0 & 0 & 0 & -t' & -\mathrm{i}t_0 \\
\end{array}
\right)
\left(
\begin{array}{c}
b\\
\phi_L\\
c\\
\phi_R\\
d\\

\end{array}
\right)
=
\left(
\begin{array}{c}
-\mathrm{i}t_0\\
t'\\
0\\
0\\
0\\
\end{array}
\right)
,
\label{system_of_eq_1_dot}
\end{equation}
where $\phi_{L}$ ($\phi_R$) is the wavefunction on the left (right) quantum dot.
Solving for $c$ and $d$ yields the transmission functions
\begin{equation}
T_{L\rightarrow C}(E)=\frac{\gamma^2[(E-v_g)^2+\gamma^2/4]}{(E^2-v_g^2-3\gamma^2/4)^2+4\gamma^2E^2}
\label{trans_L_C_1_dot}
,
\end{equation}
\begin{equation}
T_{L\rightarrow R}(E)=\frac{\gamma^4}{(E^2-v_g^2-3\gamma^2/4)^2+4\gamma^2E^2}
;
\label{trans_L_R_1_dot}
\end{equation}
as above, $\gamma=2t'^2/t_0$. 
Due to symmetry, the transmission function for tunneling from the right lead to the center is given by $T_{R\rightarrow C}(E)=T_{L\rightarrow C}(-E)$. 
In Fig.~\ref{coherent_trans} the three different transmission functions are shown for small and large values of $\Delta E$.
For $\Delta E=2\gamma$ (a), the direct transmission $T_{R\rightarrow L}$ from the left to the right lead is small, but not negligible, whereas for $\Delta E=10\gamma$ (b) it is close to zero. 
At the same time, $T_{L\rightarrow C}$ and $T_{R\rightarrow C}$ show a pronounced asymmetry for small $\Delta E$, while there are nearly indistinguishable from a Lorentzian of width $\gamma$ centered at $\mp \Delta E /2$ for larger $\Delta E$. 

\begin{figure}[ht]
  \includegraphics[width=0.47\textwidth]{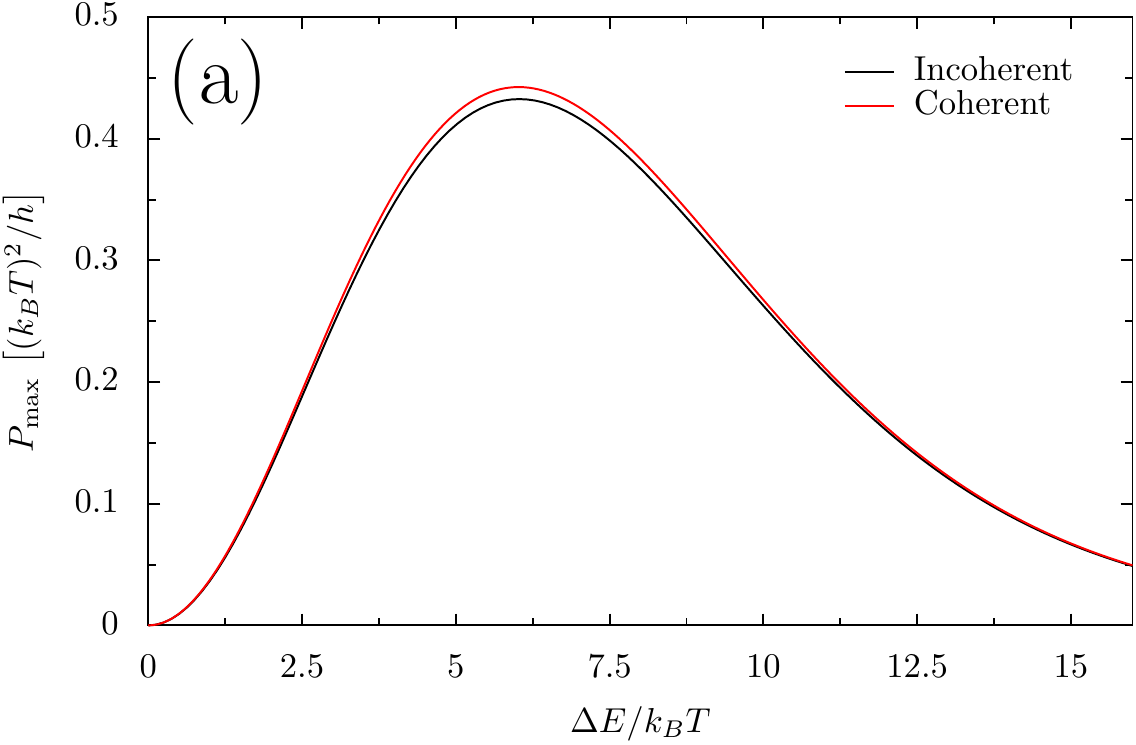}
  \includegraphics[width=0.47\textwidth]{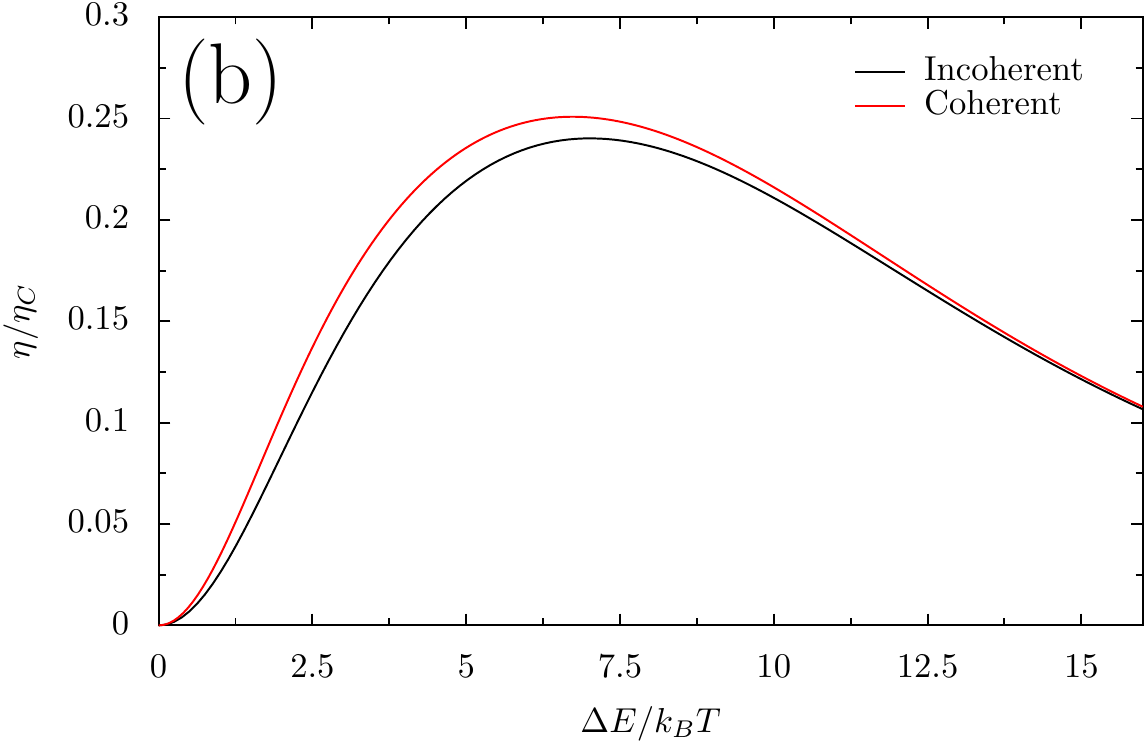}
	\caption{Maximum power (a) and efficiency at maximum power (b) of the coherent and the incoherent quantum dot heat engine as a function of $\Delta E$, for $\Delta T/T = 1$ and $\gamma=k_B T$}
	\label{coherent_vs_incoherent_1_dot}
\end{figure}

In Fig.~\ref{coherent_vs_incoherent_1_dot} the maximum power and the efficiency at maximum power of both models (coherent and incoherent) are shown for $\gamma=k_B T$ and $\Delta T/T = 1$.
The coherent heat engine performs better in the whole parameter range, but the benefit compared to the incoherent device is only marginal. 
This is due to the fact that at maximum power the working point of the engine is such that the direct coherent transfer of electrons from the left to the right lead is strongly suppressed.  

\section{Optimized transmission functions}
\label{sec_opt_trans}

As already pointed out in Sec.~\ref{introduction}, the optimal performance of the heat engine is achieved for a transmission function of rectangular shape due to its perfect energy filtering property. \citep{whitney2014,whitney2015}
Quantum dots with a Lorentzian transmission function are clearly suboptimal in this respect.
In the following, we investigate alternative realizations based on small arrays of coupled quantum dots, in order to get as close as possible to the optimum. 
In particular, we analytically calculate the transmission functions of homogeneous linear arrays (see Fig.~\ref{linear_chain}) and rings 
(see Fig.~\ref{ring_structure}) of arbitrary size, and derive the  optimal hopping parameters of inhomogeneous linear arrays of up to 
13 coupled quantum dots.

\subsection{Homogeneous linear chain}
\label{sec_homo_lin_chain}

\begin{figure}[ht]
  \includegraphics[width=0.6\textwidth]{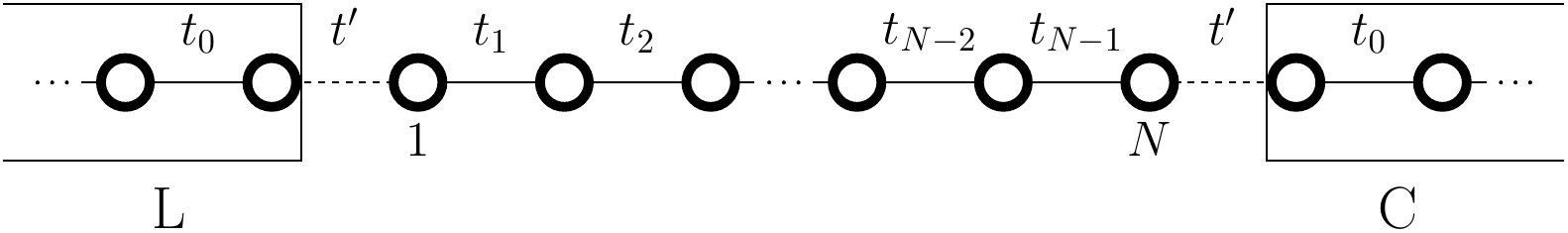}
	\caption{Linear chain with arbirary hopping parameters $t_n$ coupled to one-dimensional leads}
	\label{linear_chain}
\end{figure}
As a first example, we consider a linear array of $N$ quantum dots coupled to left and right leads as shown in Fig.~\ref{linear_chain}.
We assume that the hopping parameters within the array are all identical, $t_n \equiv t_C$.
Similar systems have previously been studied \citep{mardaani} using the Green's function formalism,\citep{datta} in particular, the Caroli formula\citep{caroli} $T(E) = \mathrm{Tr}(\Gamma G^R \Gamma G^A)$, exploiting certain recurrence relations for the Green's function for chains of different lengths. 
We find it simpler and more transparent to derive the transmission function by explicitly solving the one-dimensional scattering problem. 
Within the quantum dot array, the general solution of the discrete Schr\"odinger equation at energy $E = -2t_C \cos q $ is given by
\begin{equation}
\phi_n = A \mathrm{e}^{\mathrm{i}qn} + B \mathrm{e}^{-\mathrm{i}qn} 
\label{eigenstates_lin}
\end{equation}
for $1\le n\le N$. 
For given $\phi_1$ and $\phi_N$ the coefficients $A$ and $B$ can be determined, and thus the solution of the Schr\"odinger equation within the array is explicitly known.
In order to calculate the transmission function, we again employ the ansatz of an incoming plane wave with energy $E = -2t_0 \cos k$ that is reflected with amplitude $b$ and transmitted with amplitude $c$. 
Inserting this ansatz into the Schr\"odinger equation for the whole system yields the matrix equation
\begin{equation}
\left(
\begin{array}{cccc}
 t_0\mathrm{e}^{-\mathrm{i}k}& -t'& 0 & 0  \\
-t' & \frac{\sin Nq}{\sin (N-1)q}t_C & -\frac{\sin q}{\sin (N-1)q} t_C & 0 \\
0 & -\frac{\sin q}{\sin (N-1)q} t_C &\frac{\sin Nq}{\sin (N-1)q}t_C &  -t' \\
 0 & 0 & -t' & t_0\mathrm{e}^{-\mathrm{i}k}  \\
\end{array}
\right)
\left(
\begin{array}{c}
b\\
\phi_1\\
\phi_N\\
c\\
\end{array}
\right)
=
\left(
\begin{array}{c}
-t_0\mathrm{e}^{\mathrm{i}k}\\
t'\\
0\\
0\\
\end{array}
\right)
\label{sys_eq_lin_chain1}
\end{equation}
for the four unknowns $\phi_1$, $\phi_N$, $b$, and $c$.
Solving for $c$ yields the transmission function
\begin{equation}
T(E)=|c|^2=\left|\frac{2\alpha \sin k\, \sin q}{\mathrm{e}^{-\mathrm{i}k}\sin (N+1)q-2\alpha\sin Nq+\alpha^2\mathrm{e}^{\mathrm{i}k}\sin (N-1)q}\right|^2
,
\label{lin_trans}
\end{equation}
with $\alpha = t'^2/t_Ct_0$.
In the wide-band limit ($k\rightarrow \pi/2$) and for the special case $\alpha=1$, we obtain the following particularly simple result: 
\begin{equation}
T(E)=\frac{1}{1+\cot^2q\,\sin^2Nq}
\label{lin_trans_simple}
.
\end{equation}
Equations (\ref{lin_trans}) and (\ref{lin_trans_simple}) are only valid inside the energy band of the linear array, i.e., for $-2t_C < E < 2 t_C$.
It is, however, straightforward to generalize these results for energies outside this energy range, where the solution of the Schr\"odinger equation 
is given by 
\begin{equation}
\phi_n =
\left\{
\begin{array}{ll}
A \mathrm{e}^{qn} + B \mathrm{e}^{-qn} &\mathrm{for}\ E < -2t_C,\\
(-1)^n (A \mathrm{e}^{qn} + B \mathrm{e}^{-qn}) &\mathrm{for}\ E > 2t_C;\\
\end{array}
\right.
\end{equation}
\begin{figure}[ht]
	\includegraphics[width=0.6\textwidth]{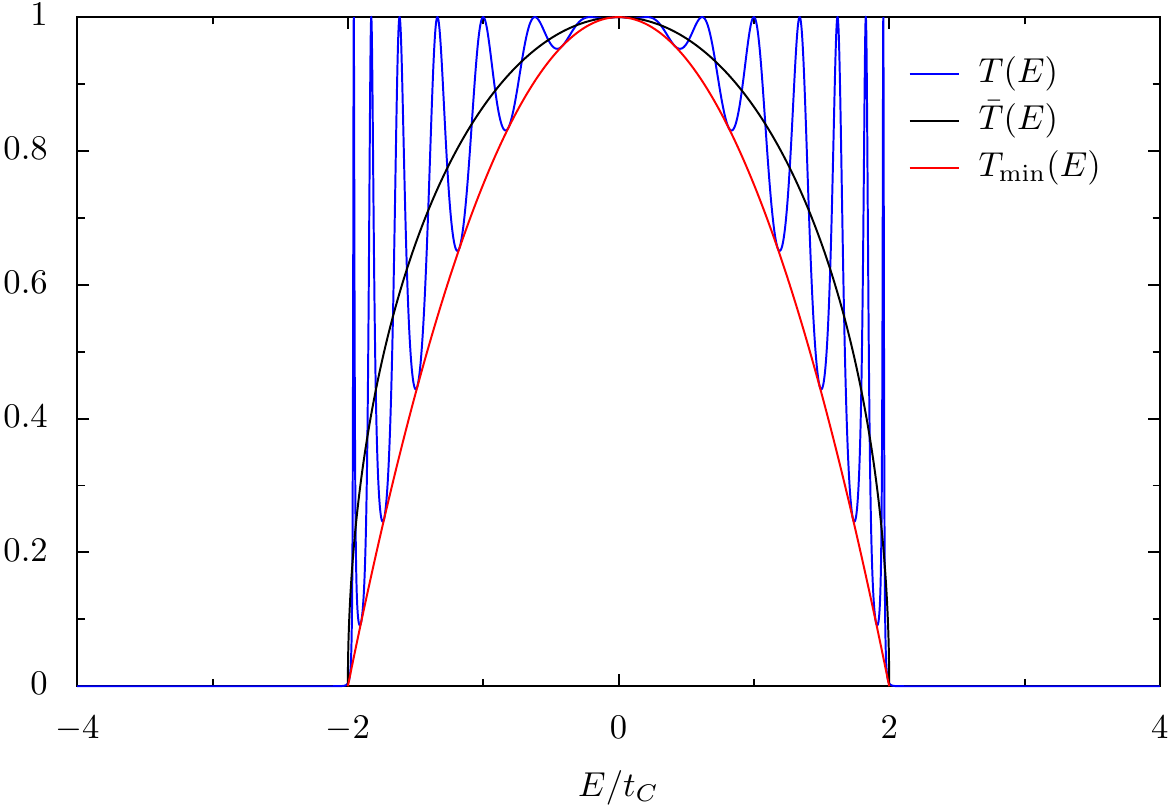}
	\caption{$T(E)$ for the homogeneous linear chain in the wide-band limit, for $\alpha=1$ and $N=15$. The lower bound $T_{\text{min}}$ and
		the averaged transmission function $\bar T$ are also shown.}
  	\label{trans_lin_chain_same_tc}
\end{figure}
here the energy is parametrized as $|E| = 2t_C \cosh q$.
As a result, for energies $|E| > 2t_C$ one has to substitute $\sin(nq) \rightarrow \sinh(nq)$, $\alpha\rightarrow\alpha\,\text{sign}(-E)$, 
and $\cot q \rightarrow \coth q$ in Eqs.~(\ref{lin_trans}) and (\ref{lin_trans_simple}), respectively.
Due to the hyperbolic functions in the denominator, $T(E)$ vanishes exponentially for $|E| > 2t_C$, especially fast for large $N$. 
In the range $|E| < 2t_C$ the transmission function oscillates between maxima with perfect transmission, and minima that are bounded from below by 
\begin{equation}
T_{\text{min}}(E) = 1 - \left(\frac{E}{2t_C}\right)^2
.
\end{equation}
The maxima are located at $E=0$ and $E_\nu = -2t_C \cos q_\nu$, with $q_\nu = \nu\pi/N$ and $1 \le \nu \le N-1$. 
In the large-$N$ limit one may average over the rapid oscillations, and obtains the semicircular transmission function
\begin{equation}
\bar T(E) = \int_0^{2\pi} \frac{d\varphi}{2\pi} \frac{1}{1+\cot^2q \sin^2\varphi} = \sqrt{1 - \left(\frac{E}{2t_C}\right)^2}
\label{average_trans}
\end{equation}
which is a good approximation when used in the energy integrals of the Landauer formula instead of the exact $T(E)$. 
In Fig.~\ref{trans_lin_chain_same_tc} we show $T(E)$ for a chain of $N = 15$ quantum dots, together with the lower bound $T_{\text{min}}(E)$ and the
averaged transmission $\bar T(E)$.
The transmission function of a homogeneous linear chain is clearly better than the Lorentzian of a single quantum dot, 
in particular, for large $N$, but still deviates substantially from the ideal rectangular profile we are aiming at.

\subsection{Homogeneous ring structure}

\begin{figure}[ht]
  \includegraphics[width=0.5\textwidth]{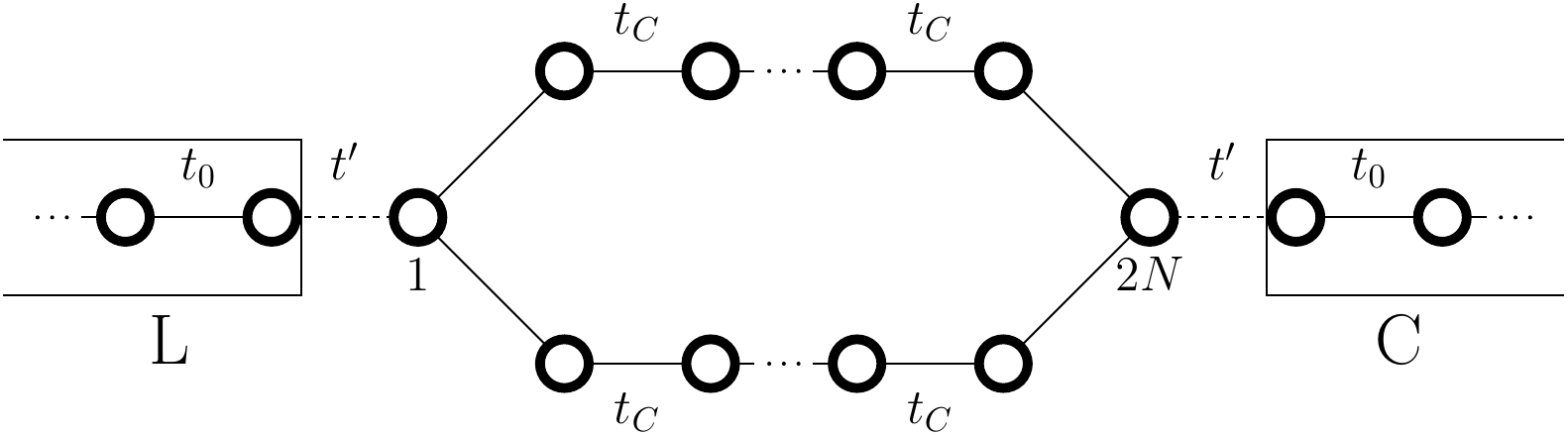}
	\caption{Homogeneous ring structure attached to one-dimensional leads}
	\label{ring_structure}
\end{figure}

As a second example, we consider a ring structure of $2N$ quantum dots symmetrically attached to the leads as shown in Fig.~\ref{ring_structure}.
In both arms, upper and lower, the general solution of the Schr\"odinger equation for the energy $E = -2t_C \cos q$ is given in the form of 
Eq.~(\ref{eigenstates_lin}).
Thus, for given boundary values $\phi_1$ and $\phi_{2N}$, the wavefunction is known for all sites within the ring. 
The scattering problem for an incoming wave with wavenumber $k$ and energy $E = - 2t_0 \cos k$ leads to the linear system   
\begin{equation}
\left(
\begin{array}{cccc}
 t_0\mathrm{e}^{-\mathrm{i}k}& -t'& 0 & 0  \\
-t' & 2t_C\frac{\sin q\cos Nq}{\sin Nq} & -2t_C\frac{\sin q}{\sin Nq}  & 0 \\
 0 & -2t_C\frac{\sin q}{\sin Nq}& 2t_C\frac{\sin q\cos Nq}{\sin Nq}& -t' \\
 0 & 0 & -t' & t_0\mathrm{e}^{-\mathrm{i}k}  \\
\end{array}
\right)
\left(
\begin{array}{c}
b\\
\phi_{1}\\
\phi_{2N}\\
c\\
\end{array}
\right)
=
\left(
\begin{array}{c}
-t_0\mathrm{e}^{\mathrm{i}k}\\
t'\\
0\\
0\\
\end{array}
\right)
.
\label{sys_eq_lin_chain2}
\end{equation}
\begin{figure}[ht]
  \includegraphics[width=0.6\textwidth]{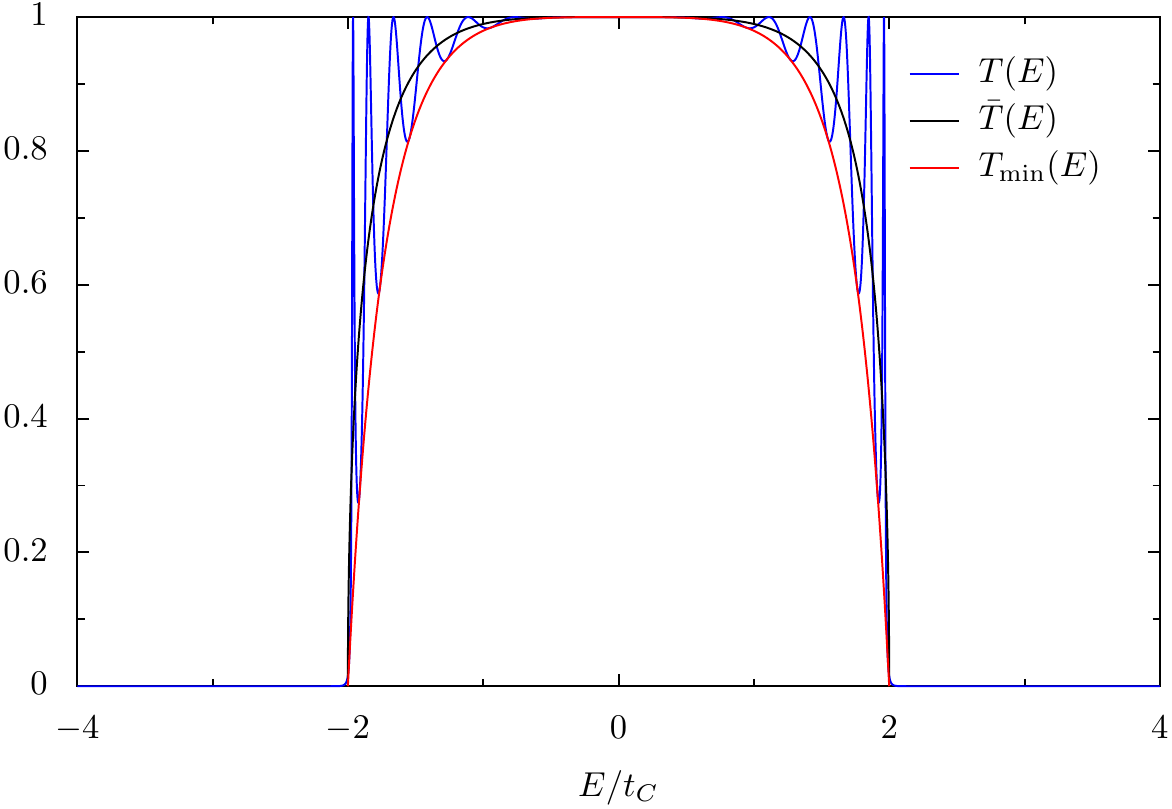}
	\caption{
	Transmission function of the ring structure (see Fig.~\ref{ring_structure}) in the wide-band limit, for $\alpha=2$ and $N=15$.
	The lower bound $T_{\text{min}}$ from Eq.~(\ref{Tmin_ring}) and the averaged transmission function $\bar T$ 
	from Eq.~(\ref{average_trans_ring}) are shown for comparison. 
	}
	\label{trans_ring}
\end{figure}
Solving for $c$ yields the transmission function
\begin{equation}
T(E)=\left|\frac{4\alpha \sin k\, \sin q\, \sin Nq}{\left(2\sin q\,\cos Nq-\alpha\mathrm{e}^{\mathrm{i}k}\sin Nq\right)^2-4\sin^2q}\right|^2
;
\label{trans_ring_eq}
\end{equation}
as above, $\alpha=t'^2/t_0t_C$.
In the wide-band limit ($k\rightarrow \pi/2$), and for the special value $\alpha = 2$, Eq.~(\ref{trans_ring_eq}) simplifies to 
\begin{equation}
T(E)=\frac{4}{4 + (\cot q\,\cos q\,\sin Nq)^2}
,
\end{equation}
which is easier to analyze than the general expression (\ref{trans_ring_eq}). 
Figure \ref{trans_ring} shows $T(E)$ for rings with $10$ and $20$ dots.
Similar to the linear chain, $T(E)$ has maxima at $E_\nu = -2t_C \cos q_\nu$ with $q_\nu = \nu\pi/N$, but now the lower
bound is given by 
\begin{equation}
T_{\text{min}}(E) = \frac{4(1 - \epsilon^2)}{(2 - \epsilon^2)^2} 
\label{Tmin_ring}
\end{equation}
with $\epsilon = E/2t_C$. 
Integrating over the rapid oscillations in the limit $N \gg 1$ yields  
\begin{equation}
\bar T(E) = \frac{\sqrt{1 - \epsilon^2}}{1 - \frac{\epsilon^2}{2}} = 1 - \frac{\epsilon^4}{8} + {\cal O}(\epsilon^6),
\label{average_trans_ring}
\end{equation}
which is superior to the semicircular averaged transmission function, Eq.~(\ref{average_trans}), with regard to the performance of the heat engine.

\subsection{Inhomogeneous linear chain}

In order to further improve the transmission function of the homogeneous linear chain, 
we now consider a linear array with arbitrary hopping parameters $t_n$ as depicted in Fig.~\ref{linear_chain}.
This problem has been addressed before in Ref.~\onlinecite{whitney2015}, using a trial and error method for small chains.
Here, we pursue a more rigorous approach.
In the wide-band limit, the Schr\"odinger equation for the corresponding scattering problem reads
\begin{equation}
\left(
\begin{array}{ccccccc}
-\mathrm{i}t_0 & -t' & & & & & \\
-t' & -E &-t_1 & & & & \\
 & -t_1 & -E & -t_2 & & & \\
 & & \ddots & \ddots& \ddots& & \\
 & & & -t_{N-2} & -E &-t_{N-1} & \\
 & & & & -t_{N-1} & -E & -t' \\
 & & & & & -t'& -\mathrm{i}t_0 \\
\end{array}
\right)
\left(
\begin{array}{c}
b\\
\phi_1\\
\phi_2\\
\vdots\\
\phi_{N-1}\\
\phi_N\\
c\\
\end{array}
\right)
=
\left(
\begin{array}{c}
-\mathrm{i}t_0\\
t'\\
0\\
\vdots\\
0\\
0\\
0\\
\end{array}
\right)
.
\label{sys_eq_inh_chain}
\end{equation}
Solving for $c$ yields the transmission function
\begin{equation}
T(E) = |c|^2 = \left|\frac{2t_0 t'^2 \prod_{n=1}^{N-1}t_n}{D(E)}\right|^2
\label{trans_inh_chain}
\end{equation}
where $D(E)$ is the determinant of the matrix on the l.h.s.\ of Eq.~(\ref{sys_eq_inh_chain}). 
For strictly one-dimensional systems without loops, one may absorb the sign of the hopping parameters in the definition of the local states of the chain without affecting $T(E)$.
Inverting the sign of all hopping parameters and simultaneously $E$ leaves Eq.~(\ref{sys_eq_inh_chain}) invariant; thus $T(E) = T(-E)$.
In addition, $|D(E)|^2$ is an even polynomial of order $2N$. In the following, we assume $t_n > 0$, for definiteness.
Expressing $t_n$ in units of $t_C = t'^2 / t_0$ as $t_n = a_n t_C$, one obtains explicitly
\begin{equation}
T(E) = \frac{1}{c_0 + c_1\epsilon^{2} + \ldots + c_{N-1} \epsilon^{2N-2} + \epsilon^{2N}},
\label{trans_inh_chain_expl}
\end{equation}
with $\epsilon = E/\gamma$ and $\gamma = (2\prod_{n=1}^{N-1} a_n)^{1/N} t_C $.
For $E=0$ it is straightforward to evaluate the determinant $D(0)$ in Eq.~(\ref{trans_inh_chain}) for arbitrary parameters $a_n$, and from there the coefficient $c_0$. 
The result is
\begin{equation}
c_0 = \left(\frac{p_{\text e}^2 + p_{\text o}^2}{2p_{\text e}p_{\text o}}\right)^2
\end{equation}
with 
\begin{equation}
p_{\text e} = \prod_{n\ \text{even}} a_n, \ \ p_{\text o} = \prod_{n\ \text{odd}} a_n .
\end{equation}
\begin{table}[h]
	\begin{tabular}{| l | c | c | c | c | c | c |}
	\hline
	\multicolumn{1}{|c|}{$N$}  & $a_1$ & $a_2$ & $a_3$ & $a_4$ & $a_5$ & $a_6$ \\ \hline
	$\phantom{1}2$ & $1$  &  &  &  &  &  \\ \hline
	$\phantom{1}3$ & $1/\sqrt{2}$  &  &  &  &  &  \\ \hline
	$\phantom{1}4$ & $\sqrt{\sqrt{2} - 1}$  & {$\sqrt{2}-1$}  &  &  &  &  \\ \hline
	$\phantom{1}5$ & {$\sqrt{5}/2  - 1/2$} & $\sqrt{\sqrt{5}/2-1}$  &  &  &  &  \\ \hline
	$\phantom{1}6$ & $\sqrt{\sqrt{3}/2-1/2}$  & $\sqrt{3\sqrt{3}/2-5/2}$  & {$2 - \sqrt{3}$} &  &  &  \\ \hline
	$\phantom{1}7$ & 0.597408  & 0.296896  & {0.234432}   &  &  &\\ \hline
	$\phantom{1}8$ & $0.593$ & $0.287$ & $0.216$  & $0.199$ &  & \\ \hline
	$\phantom{1}9$ & $0.589$ & $0.281$ & $0.205$  & $0.179$ &  & \\ \hline
	$10$ & $0.587$ & $0.276$ & $0.197$  & $0.167$ & $0.158$  & \\ \hline
	$11$ & $0.585$ & $0.273$ & $0.192$  & $0.158$ & $0.145$  & \\ \hline
	$12$ & $0.584$ & $0.270$ & $0.188$  & $0.152$ & $0.136$  & $0.132$ \\ \hline
	$13$\ \ \ \ & $0.583$ & $0.268$ & $0.185$  & $0.148$ & $0.130$  & $0.122$ \\ \hline
	\end{tabular}
	\caption{Parameters $a_n$, calculated analytically for systems of linear chains up to seven dots, and numerically for longer chains}
	\label{an_analytical}
\end{table}

Perfect transmission at zero energy, $T(0) = 1$, is only possible for $p_{\text e} = p_{\text o}$. 
In order to obtain a transmission function that resembles as far as possible the rectangular one, we have to adjust the hopping parameters such that $c_0 = 1$, and $c_1 = c_2 = ... = c_{N-1} = 0$ in Eq.~(\ref{trans_inh_chain_expl}).
Then $T(E)$ assumes the particularly simple form
\begin{equation}
T(E) = \frac{1}{1 + \epsilon^{2N}},
\end{equation}
which interpolates between the Lorentzian ($N = 1$) and the rectangular function ($N \rightarrow\infty$). 
For $N \le 6$ the solution of the coupled nonlinear equations that follow from the above conditions can be given in analytical form, while 
for larger $N$ we calculate the parameters $a_n$ numerically using Newton's method. 
In all cases there exists a unique real solution. This solution is symmetric, $a_{N-n} = a_n$, and
decreases monotonically from the boundary to the center of the chain. 
In Tab.~\ref{an_analytical} we list the coefficients $a_n$ for systems of up to $N = 13$, taking into account the symmetry property.

\begin{figure}[ht]
  \includegraphics[width=0.6\textwidth]{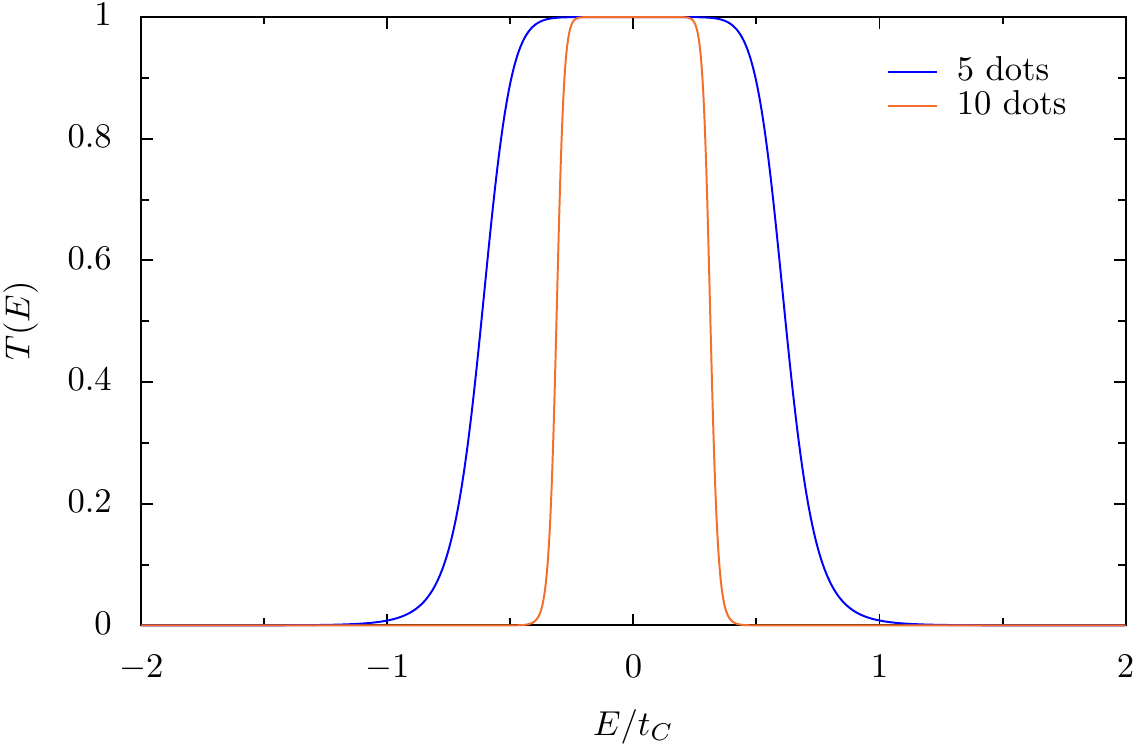}
	\caption{Transmission function of the optimized inhomogeneous chain, for $N=5$ and $N=10$}
	\label{trans_lin_chain_lorentzian}
\end{figure}

The shape of the transmission function, using the optimized hopping parameters for chains of $N=5$ and $N=10$ dots, is shown in 
Fig.~\ref{trans_lin_chain_lorentzian}.
As expected, the curve for larger $N$ is nearly rectangular, but at the same time its width $\gamma = (2\prod_{n=1}^{N-1} a_n)^{1/N} t_C$
is reduced. To compensate for this effect, one has to readjust the parameter $t_C$ accordingly.

Dissipationless resonant tunneling heterostructures similar to ours have recently also been studied in the context of space inversion and
time reversal symmetry breaking, \cite{gorbatsevich} but without considering possible applications to nano-devices.

\subsection{Comparison}

\begin{figure}[ht]
  \includegraphics[width=0.47\textwidth]{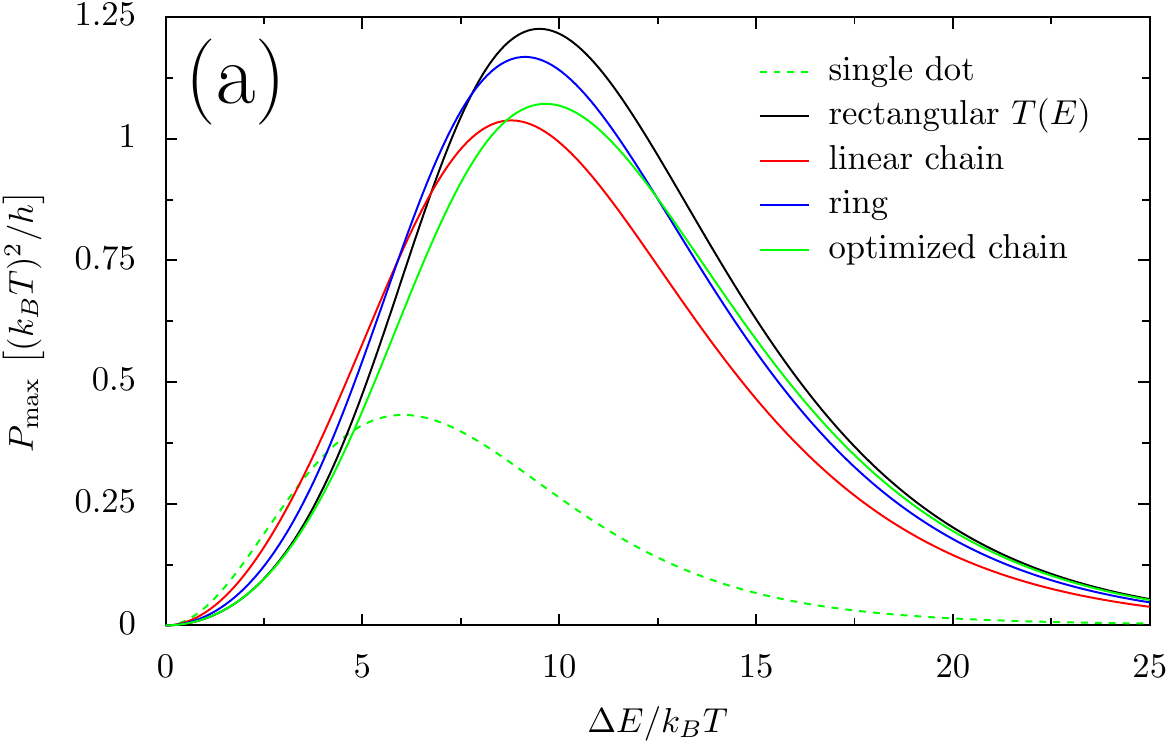}
  \includegraphics[width=0.47\textwidth]{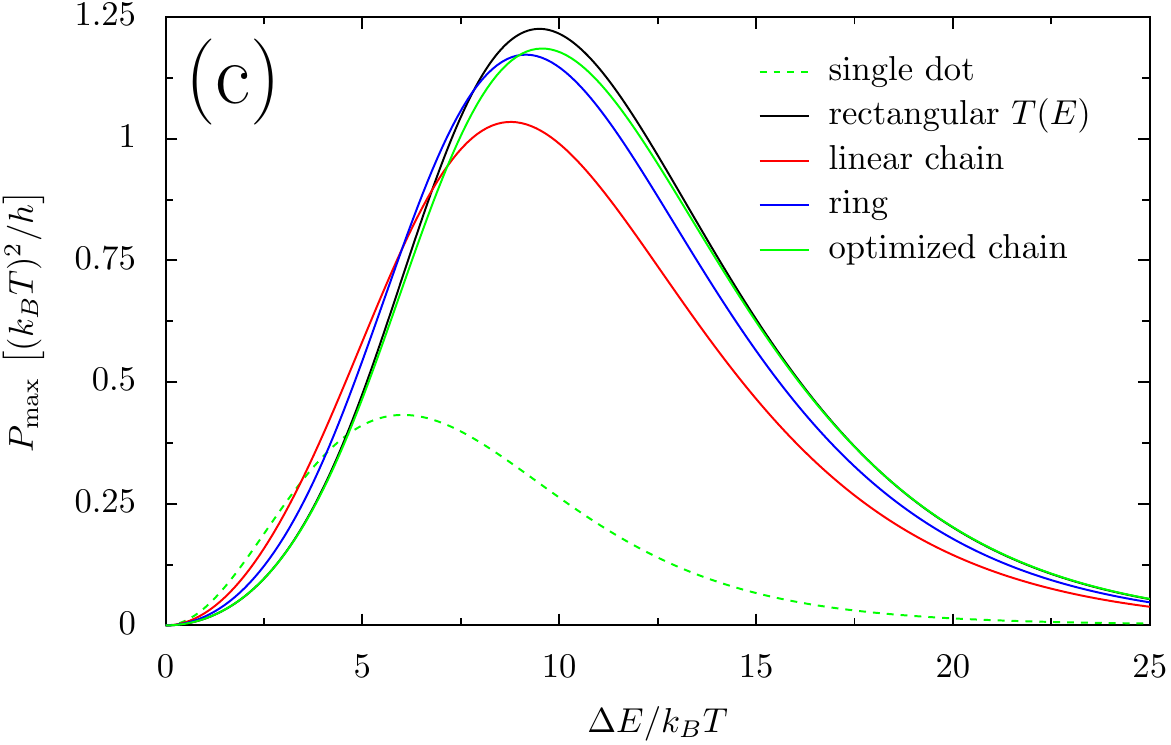} 
  \includegraphics[width=0.47\textwidth]{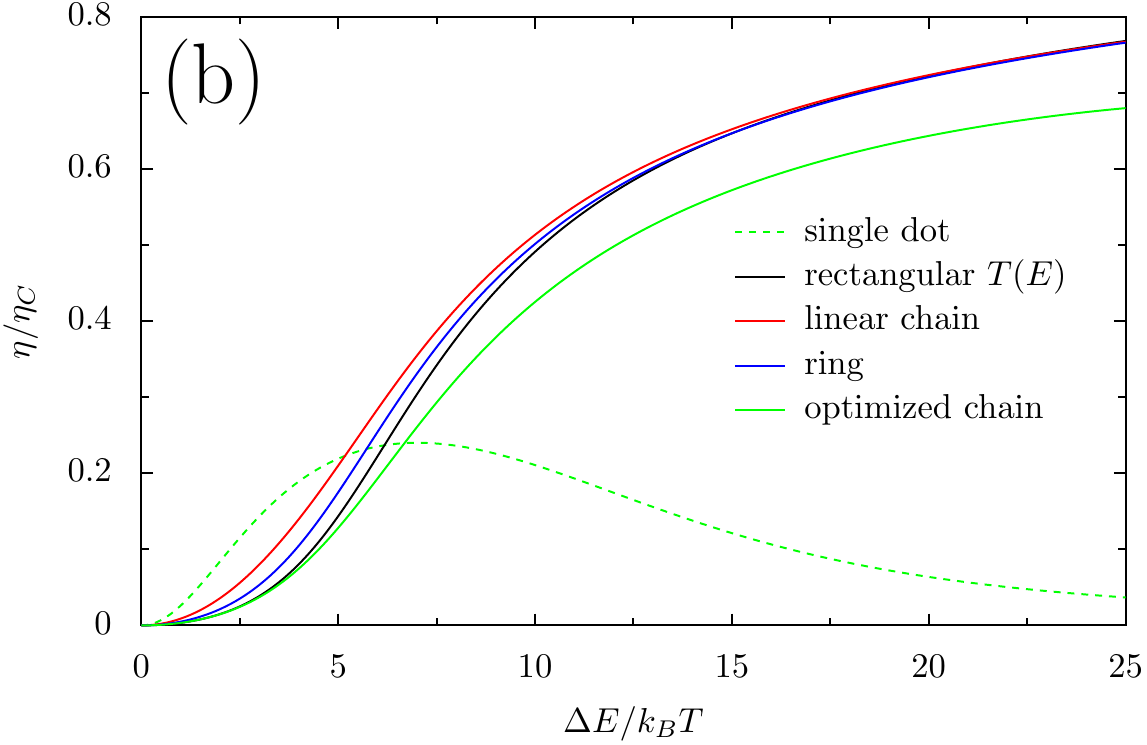}
  \includegraphics[width=0.47\textwidth]{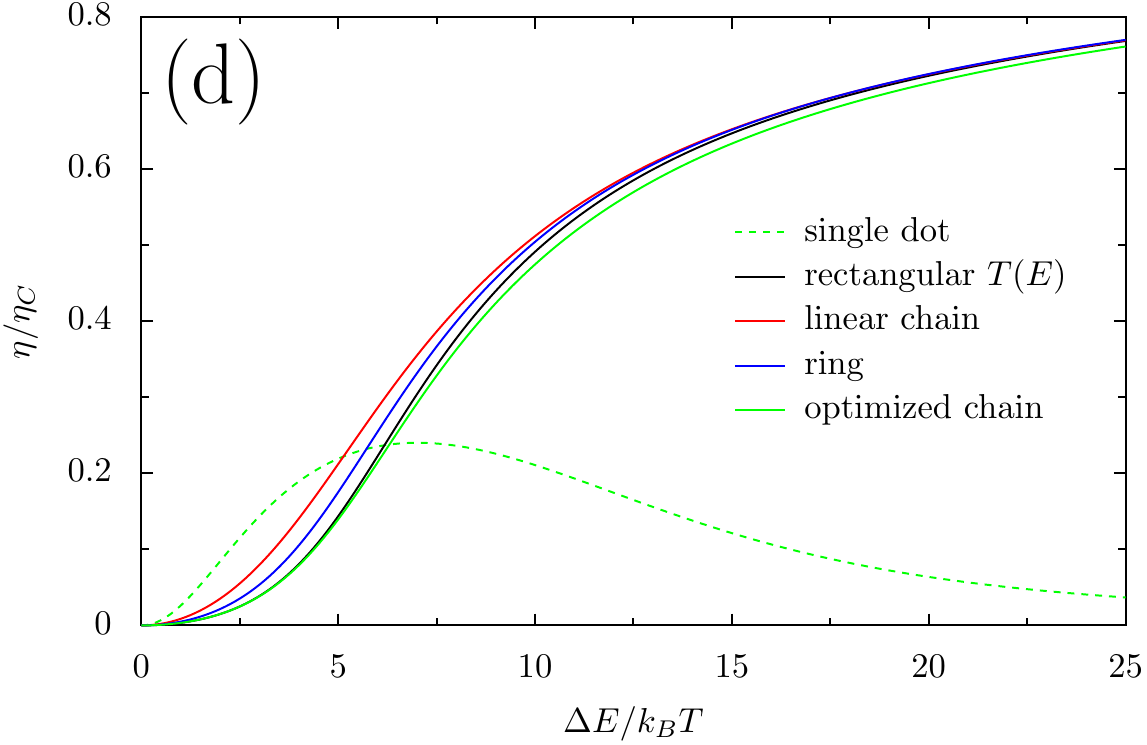}
	\caption{Maximum power and efficiency at maximum power of the incoherent heat engine for various transmission functions, 
	with parameters $\gamma = k_BT$ (single dots), $t_C = 1.5k_BT$ (homogeneous chains and rings), $\gamma = 3k_BT$ (inhomogeneous chains), 
	and $\Delta T/T = 1$. Panels (a) and (b) are for $N=5$, panels (c) and (d) for $N = 10$.
	}
	\label{power_eff_incoherent}
\end{figure}

To conclude this section, we compare the quantum dot arrays investigated above with respect to their performance as energy filters with the single quantum dots originally proposed. 
For simplicity, we consider the coherent model, since, based on our findings for the single-dot engine, we do not expect the results to be significantly different in the incoherent case. 
In order to facilitate the comparison, we choose the width of the transmission functions (parametrized by $\gamma$ for the Lorentzian and for the inhomogeneous chain, 
and by $t_C$ for the homogeneous quantum dot arrays) 
such that $P_{\text{max}}$ as function of $\Delta E$ reaches the largest possible value.
This is the case for $\gamma \approx k_BT$ (Lorentzian),  $\gamma \approx 3k_BT$ (inhomogeneous chains), and for $t_C \approx 1.5k_BT$ (homogeneous chains and rings). 
Figure \ref{power_eff_incoherent} shows the maximum power and the efficiency at maximum power of these structures for $N = 5$ and $N = 10$, in comparison
with the ideal rectangular transmission function of the same width, $w = 4t_C$.
Obviously, the performance of the heat engine can be improved significantly by replacing the single quantum dots by more appropriate multi-dot structures. 
Even the homogeneous chain of $N = 5$ dots, which is the worst among the multi-dot structures, yields an optimal $P_{\max}$ more than twice as large as of the single dot. 
Also the efficiency at the optimum of $P_{\max}$ is approximately doubled.
The performance of the heat engine can be further improved using more dots $(N=10)$ or one of the more complex structures, but the additional gain is less striking.

\section{Summary and conclusion}\label{sec_summary}

To summarize, we have analyzed two modifications of the quantum dot heat engine proposed in Ref.~\onlinecite{jordan}.
First, we have investigated a tight-binding model representing a setup where coherent tunneling between the leads is 
possible (Fig.~\ref{model_coherent}), without
intermediate thermalization, in contrast to the original proposal which corresponds to Fig.~\ref{model_incoherent}. 
The performance parameters, i.e., maximum power and efficiency at maximum power, turned out to be very similar in both cases, 
with minor advantages for the coherent setup. The explanation for this finding is the following: in order to achieve optimum 
performance the energy levels of the quantum dots (that serve as energy filters) have to be shifted in opposite directions,
such that the direct coherent transmission from the left to the right lead is strongly suppressed.

The second issue was the optimization of the heat engine by choosing arrays of tunnel-coupled quantum dots with 
more efficient transmission functions, compared to transmission through single dots. Although the ideal rectangular shape cannot be
implemented using a finite number of quantum dots, we have shown that already arrays made of very few (e.g., five) dots are
sufficient to improve both, the maximum power and the efficiency at maximum power, by a factor of two.  
We have derived simple analytical expressions for the transmission functions of homogeneous linear arrays and rings of arbitrary size,
and determined the hopping parameters for inhomogeneous linear chains such that the oscillations of $T(E)$ are completely suppressed.
The design of customized arrays of quantum dots with predefined transmission properties can be regarded as 
an inverse scattering problem. In contrast to the continuum inverse scattering problem where the potential
leading to the observed differential cross section has to be determined, here, one has to adjust the hopping parameters, 
i.e., nonlocal potentials, that yield the desired transmission. We expect that the interest in theoretical methods for such 
inverse problems will grow further, considering the ubiquitous demand for future nanoscale devices---for energy harvesting,
and other applications as well.

Finally, we note that in our approach, similar to most earlier studies of quantum dot heat engines, the Coulomb interaction among electrons, both within the dots and between neighboring dots, was not considered.
In the limit of weak correlations, it is possible to retain the Landauer picture of noninteracting particles by using decoupling schemes like the Hartree \cite{szukiewicz} or the Hartree-Fock\cite{schiegg} approximation, which roughly corresponds to taking screening into account. \cite{szukiewicz} 
Other heat engine designs, which utilize Coulomb blockade physics, have also been discussed. \cite{sothmann,sanchez,koch,thierschmann,zhang}
In order to properly treat the Coulomb blockade, or strong correlation effects in general, one has either to resort to numerical methods, or different analytical approaches. \cite{sothmann}

\acknowledgments
We are grateful to Karol I.\ Wysoki\'{n}ski and Barbara Szukiewicz for stimulating discussions. Financially supported by the German Science Foundation (DFG)
through TRR 80.

\end{document}